%% file: relics.tex
\documentstyle[ifthen,epsfig]{l-aa}
\begin{document} 
\thesaurus{ 
	    02.01.1;      
  	    02.01.2;      
  	    02.16.2;      
  	    02.19.1;      
	    11.03.1;       
	    12.12.1}       

\title{Cluster Radio Relics as a Tracer of Shock Waves of the Large-Scale
Structure Formation}
\author{Torsten A. En{\ss}lin\inst{1},
Peter L. Biermann\inst{1}, Ulrich Klein\inst{2}, Sven Kohle\inst{2}}
\institute{Max-Planck-Institut f\"{u}r Radioastronomie, Auf dem 
H\"{u}gel 69, D-53121 Bonn, Germany \and Radioastronomisches Institut
der Universit{\"a}t Bonn, Auf dem H\"{u}gel 71, D-53121 Bonn,
Germany
}
\offprints{T.A. En{\ss}lin, ensslin@mpifr-bonn.mpg.de}
\date{Received ??? , Accepted ???}  
\maketitle\markboth{En{\ss}lin et al.: Cluster Radio Relics as
a Tracer of Shock Waves of the Large-Scale Structure
Formation}{En{\ss}lin et al.: Cluster Radio Relics as a Tracer  of
Shock Waves of the Large-Scale Structure Formation}

\begin{abstract}
We present evidence for the existence of shock waves caused by the formation
of the large-scale structure. In some clusters of galaxies peripherally
located sources of extended diffuse radio emission exist, the so-called
cluster radio relics. They have steep radio spectra but no apparent cutoff, as
old remnants of radio galaxies usually have. Therefore particle acceleration
has to take place within them. We propose that shock structures of the
cosmological large-scale matter flows are responsible for the {
acceleration of} relativistic electrons: cluster accretion shocks and bow
shocks of merger events. We develop a theory of radio plasma having {
traversed} these shocks and compare it to observational data of { nine}
radio relics (0038-096, 0917+75, 1140+203, 1253+275, { 1712+64, }1706+78,
2006-56, { 2010-57,} 1401-33) and their host clusters (A85, A786, A1367,
Coma, A2255, A2256, A3667, S753). The necessary accretion power, the spectral
index of the radio spectrum, the acceleration efficiency of the shock, the
diffusion coefficient in the post-shock region, and the predicted radio
polarization in all of our examples fit into a coherent interpretation of the
observational data. Since polarization measurements are available only for
four sources, the predictions of our theory can be independently checked using
other examples. The predicted values of the shock compression ratio, density
and temperature of the infalling gas, magnetic field strength of the shocked
and unshocked radio plasma are discussed within the frame of structure
formation theory.\\
\end{abstract}
\section{Introduction}
Large extended sources of synchrotron emission are found in some clusters of
galaxies: radio halos and radio relics (Fig. \ref{coma.ps}{; for reviews,
see Jaffe 1992\nocite{jaffe92}; Feretti \& Giovannini
1996\nocite{feretti96}}). The rare cluster {\it radio halo} phenomena are {\it
large regions of diffuse unpolarized radio emission, centrally located}, and
have { radio structures} roughly similar to that of the thermal X-ray
emission { (Deiss et al. 1997)}\nocite{deiss97}.  The so-called {\it radio
relics} are {\it peripherally located, and are more irregularly shaped
polarized radio sources}. Radio relics are believed to be the remnants of
radio lobes of radio galaxies, where the former active galaxy has become
inactive or has moved away.

\ifthenelse{
\value{page}= 2 \or 
\value{page}= 4 \or
\value{page}= 6 \or
\value{page}= 8 \or
\value{page}= 10 \or
\value{page}= 12 \or
\value{page}= 14 \or
\value{page}= 16 \or
\value{page}= 18
}{\begin{figure*}[tbp]\end{figure*}}{ }

\begin{figure*}[tbh]
\psfig{figure=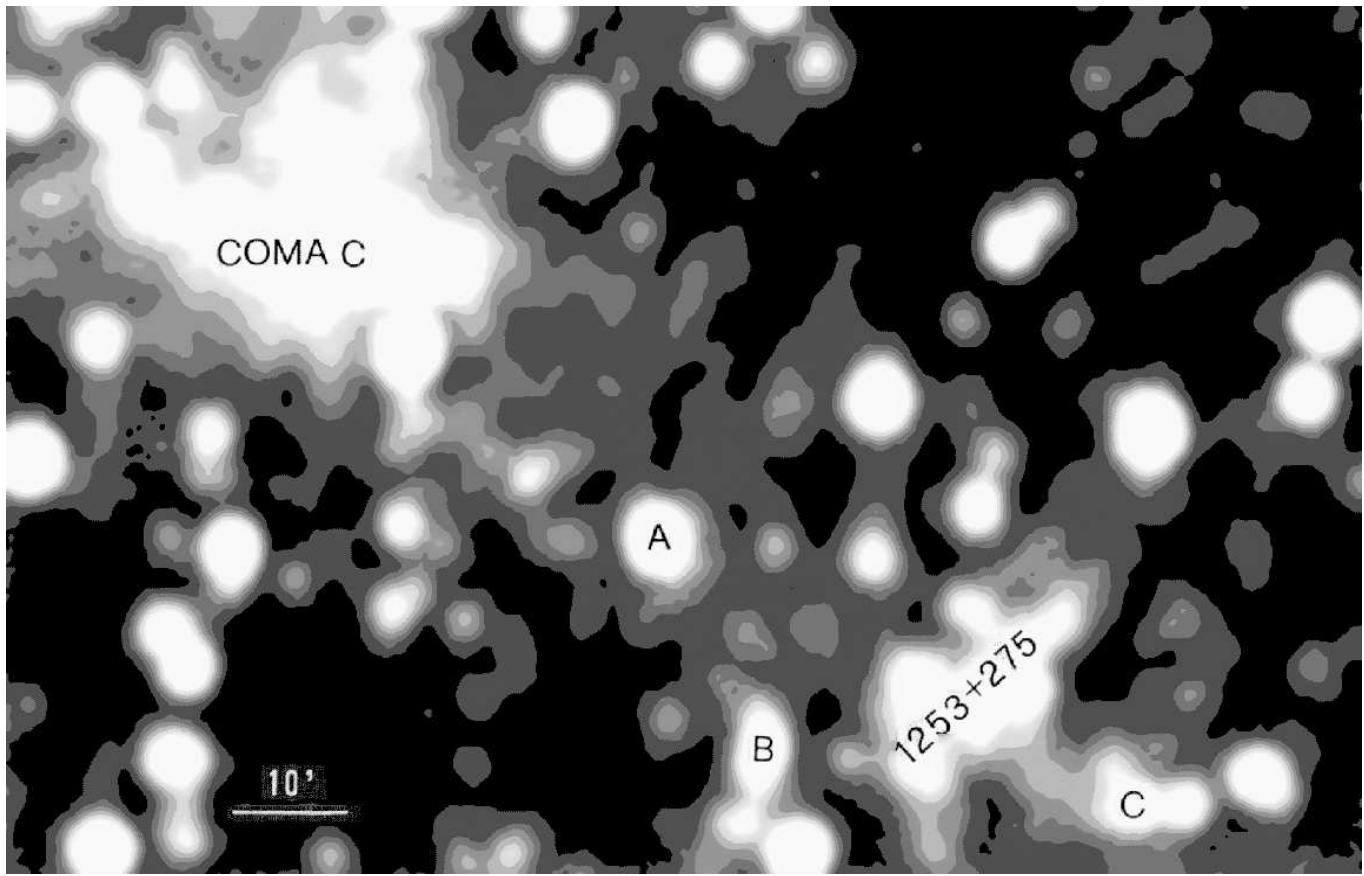,width=1.0\textwidth}
\caption[]{\label{coma.ps}Westerbork Synthesis Radio Telescope
map at 327 MHz of the Coma cluster from Giovannini et al. (1991). The
central halo source Coma C and the cluster relic 1253+275 are
identified. Capital letters indicate some extended Coma cluster galaxies,
as following: A = NGC 4839, B = NGC 4827, C = NGC 4789. 10' corresponds
to 400 kpc $h_{50}^{-1}$.}
\vspace{2em}
\end{figure*}

For radio halos a number of candidate sources for the relativistic,
synchrotron emitting electrons were discussed in the literature
(Schlickeiser et al. 1987 { and references
therein})\nocite{schlickeiser87}. Yet, the source of the energetic
electrons in the relics is unclear, due to the shortness of the
electron cooling time compared to the time since electron injection
from a galaxy at the location of the relic was possible.  Since the
radio spectra of relics are steep but frequently do not exhibit any
cutoff in the observed range, an efficient particle acceleration
mechanism has to be present within or close to them. We show that
shocks of the large-scale gas motion are expected at the typical
peripheral locations of relics, either resulting from cluster mergers,
or from steady state accretion shocks by gas falling into the cluster
potential.  Kang et al.~(1997)\nocite{kang97} showed that protons
might be accelerated in cluster accretion shocks to energies
comparable with the most energetic cosmic ray events observed at
earth. Also electrons should be accelerated and become visible by
radio emission at locations where magnetic fields are
present. Therefore the magnetized plasma accumulated behind the shock
or left behind as a remnant of a radio galaxy, could furnish this
acceleration region and should also increase the efficiency of the
acceleration mechanism.  We demonstrate that for the case of the radio
relic 1253+275 close to Coma A (Fig. \ref{coma.ps}), the observed
spectral index and polarization of the radio emission agree well with
the prediction for an accretion shock.  Eight other relics are also
briefly discussed.

We adopt for the following $\Omega_{\rm o} = 1$ and $H_{\rm o}= 50\,h_{50}\,
{\rm km\, s^{-1}\, Mpc^{-1}}$.

\section{Particle acceleration at cluster accretion shocks}
\subsection{ \label{sec:geometry}Geometry}

\begin{figure}[tbh]
\psfig{figure=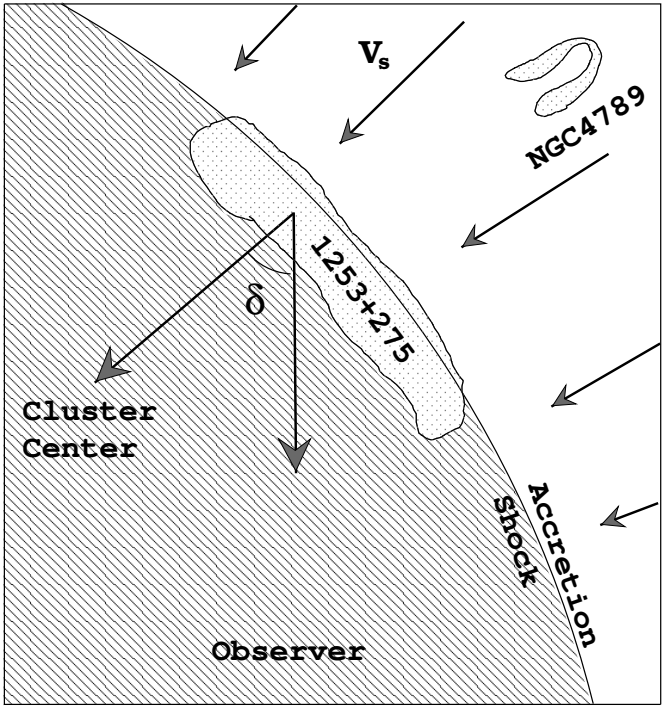,width=0.5\textwidth}
\caption[]{\label{skizze.ps}Sketch of the assumed geometry in the Coma
cluster.  The accretion flow is indicated by arrows. The shaded region is the
intra-cluster medium. The dotted area represents regions of radio plasma. The
line-of-sight and the line connecting the relic and the cluster center are
lying within the plane of the figure. NGC4789 is placed upstream above the
radio relic as a source of relativistic electrons, as discussed in
Sect. \ref{sec:origin}. Its line-of-sight velocity component points away from
the observer.  All positions could be different due to shifts parallel to the
line-of-sight.}
\end{figure}

Properties of the accretion shocks of clusters of galaxies, the shock radius
$r_{\rm s}$ and the velocity $V_{\rm s}$ of the infalling matter measured in
the shock frame, are related to the depth of the gravitational potential of the
cluster. Since the cluster gas temperature also depends on the potential, the
shock properties can be expressed in terms of the observed temperature $kT_{\rm
obs}$. Using results of one-dimensional simulations of accreting flows onto
clusters { (Ryu \& Kang 1997)\nocite{ryu97}}, Kang et al.~
(1997)\nocite{kang97} derived
\begin{eqnarray}
\label{eq:rs}
r_{\rm s} &=& 4.24\, h_{50}^{-1} {\rm Mpc} \left( \frac{kT_{\rm
obs}}{6.06\,{\rm keV}} \right)^{1/2} (1+z)^{-3/2}\\
\label{eq:Vs}
V_{\rm s} &=& 1750 \, {\rm km\, s^{-1}}\,\left( \frac{kT_{\rm
obs}}{6.06\,{\rm keV}} \right)^{1/2}\,\,.
\end{eqnarray}
{ These estimates depend on the assumption of a spherical flow, of a
polytropic index of the gas of $\gamma_{\rm gas} = 5/3$, and they depend
weakly on the assumed Einstein-de Sitter cosmology. Since the real infall
pattern to clusters should be aspherical and correlated with the surrounding
large-scale structure (Colberg et al. 1997)\nocite{colberg97}, the shock radii
and the shock velocities are expected to differ from these estimates. They
might be regarded as average quantities, where from a single cluster can
easily deviate.} Inserting the X-ray temperature of the Coma cluster of 8.2
keV (Briel et al.~1992)\nocite{briel92} and the redshift of $z= 0.0233$
(Fadda et al.~1996)\nocite{fadda96} gives a shock radius of $r_{\rm
s}= 4.8\, h_{50}^{-1}\,{\rm Mpc}$ and a shock velocity of $2.0\cdot 10^3
\,{\rm km
\,s^{-1}}$. If 1253+275 is at the position of this shock, the projected
distance to the cluster center of $2.9 \,h_{50}^{-1}\,{\rm Mpc}$ (Giovannini et
al.~1991)\nocite{giovannini91} implies an angle between line-of-sight and
normal of the shock front of $\delta = 37^\circ$. { The true angle is
ambiguous ($180^\circ - \delta$ or $\delta$), and depends on the location of
the relic on the line-of-sight (foreground or background with respect to the
cluster center).} A possible configuration is sketched in
Fig. \ref{skizze.ps}{ , where the relic is located on the back side of the
cluster accretion shock, since then a physical connection to the nearby radio
galaxy NGC 4789 is possible as explained in Sect. \ref{sec:origin}. The length
of the relic is $1.2\,{\rm Mpc}\,h_{50}^{-1}$, whereas the thickness cannot be
given directly from radio observations. We assume a thickness of $D = 0.1\,{\rm
Mpc}\,h_{50}^{-1}$, which is smaller than the observed projected thickness due
to the action of shock compression in the infall direction. A detailed model of
the structure of the relic would allow to deproject the apparent surface. For
only moderately large projection angles ($\delta <60^\circ$) resulting
deprojected surfaces are not much bigger than the observed ones.  Since the
geometric quantities are only used for order of magnitude calculations of the
infalling kinetic energy flux, the shock efficiency, and the diffusion
coefficient within the radio plasma, we prefer not to specify a detailed
geometrical model. We equate projected to real surface of the relic, which is a
weak underestimate. The relic is slightly V-shaped and the radio emission is
more sharply edged on the outer side (Giovannini et
al.~1991)\nocite{giovannini91}. The V-shape could result from an aspherical
shock, but would also be easily explained by an intrinsic shape of the
magnetized plasma, since the relic is not seen edge-on.  The edges of the
observed projected radio structure should differ. The radio plasma seen at the
inner edge (left edge in Fig. \ref{skizze.ps}) is more distant to the shock
than that at the outer edge (right). The outer edge should be sharper due to
the better confined plasma at the shock side, and it should have a lower than
average spectral index, since the electrons there are reaccelerated
recently. Whereas the inner edge is expected to be smoother and to have a
higher than average spectral index. The reason for the latter is that the
reaccelerated electrons within the radio plasma seen there, at a position more
distant to the shock, had more time for cooling.  }
\subsection{\label{compr}Compression ratio}
The radio spectral index $\alpha$ ($S_\nu \sim \nu^{-\alpha}$) of 1253+275 is
$\alpha= 1.18\pm0.06$ (Giovannini et al.~1991)\nocite{giovannini91} and is
therefore larger than $\alpha= 1$, expected for a strong shock accelerating
electrons, with simultaneous synchrotron and inverse-Compton losses. Larger
spectral indices are found { at the inner edge} of the relic, { showing
evidence for} steepening of the electron momentum distribution due to { the
higher age of the electrons seen there, more distant to the acceleration
region of the shock.}  The radius of the accretion shock structure (assumed to
have a spherical surface) is large compared to every intrinsic length scale of
the acceleration process. Thus the theory of planar shocks can be applied.
The spectral index can be explained by allowing the shock compression ratio
$R$ to be lower than the canonical value $R=4$ of a strong shock of gas with a
polytropic index of $\gamma_{\rm gas} = 5/3$. The radio spectral index
$\alpha$ of an equilibrium electron population accelerated and cooled at the
same time, is related to the shock compression ratio by (e.g. Drury
1983)\nocite{drury83}:
\begin{equation} \label{rs}
R = \frac{\alpha+1}{\alpha-\frac{1}{2}}\,\, ,
\end{equation}
where a polytropic index of $\gamma_{\rm gas} = 5/3$ is assumed. The
compression ratio is therefore $R=3.2\pm0.2$.  From the theory of shocks
(Landau \& Lifschitz 1966\nocite{landau66}) the pressure and temperature ratio
between the down- and up-stream region (inside and outside the cluster shock
front, in the following denoted by $P_2$ and $P_1$, etc.) can be derived:
\begin{equation} \label{P}
\frac{P_{2}}{P_{1}} = \frac{4R-1}{4-R}= \frac{\alpha
+\frac{3}{2}}{\alpha-1}\,\,\,\,\,\, , \,\,\,\,
\frac{T_{2}}{T_{1}} = \frac{1}{R} \, \frac{P_{2}}{P_{1}}  \,\,.
\end{equation}
For 1253+275 the resulting pressure ratio is $P_2/P_1= 11 - 22$. The
temperature ratio $T_{2}/T_{1} = 3.8 - 6.4$ requires that the infalling
matter has a temperature of { $T_{1} \sim 0.6 - 1.1$ keV}, assuming
the post-shock temperature to be roughly half the central cluster
temperature. Simulations predict this radial temperature decrease (Navarro et
al.~1995)\nocite{navarro95}, and observations indicate a radially falling
temperature profile in clusters (Honda et al.~1996; Markevitch 1996;
Markevitch et al.~1996,
1998)\nocite{honda96}\nocite{markevitch96}\nocite{markevitch96a}.

This temperature of the infalling matter is reasonable, since this gas should
flow mainly out of sheets and filaments of the cosmological large-scale
structure, and therefore was preheated by the accretion shocks at the
boundaries of these structures. A simulation of cosmological structure
formation by Kang et al.~(1996)\nocite{kang96} shows that typical temperatures
of filaments are above 0.1 keV. Adiabatic compression and internal shocks
within the flows onto clusters might raise these temperatures to a level of
$0.5 -  1$ keV, which is needed in order to explain the steepness of the
synchrotron spectrum of cluster relic sources caused by weak shocks.

The temperature of the infalling matter gives the sound velocity $c_{1}$, which
enters into a second estimate of the shock velocity, using the theory of
planar shock waves (Landau \& Lifschitz 1966):
\begin{equation}
\label{eq:Vsp}
V_{\rm s, predicted} = c_1\, \sqrt{\frac{(\gamma_{\rm gas}-1)+(\gamma_{\rm
gas} +1)P_2/P_1}{2 \gamma_{\rm gas}}}\,\,.
\end{equation}
This estimate gives a shock velocity for Coma of $V_{\rm s, predicted} =
1850\,{\rm km/s}$, slightly smaller than that following from the theory of
Kang et al. (1997). The difference vanishes if a smaller temperature drop of
$\approx 1.6$ from the center to the shock radius is used instead of a factor
of 2. Comparing both estimates shows self-consistency of our model, but this is
not a completely independent test, since both values depend mainly on the same
quantity ($V_{\rm s} \sim (kT_{\rm obs})^{1/2}$).

In the model of acceleration of ultra-high-energy cosmic rays at cluster
accretion shocks of Kang et al.~(1997) a compression ratio close to $R=4$ is
favored in one particular model to fit the shape of the spectrum of
ultra-high-energy cosmic rays using a low and energy independent escape
efficiency of cosmic rays from the cluster against the upstream flow. However,
a lower compression ratio (and therefore a steeper cosmic ray production
spectrum) can be compensated by taking a reasonable energy dependence of the
escape efficiency into account.  In fact, in Kang et al. (1997) the
acceleration and escape efficiency was rather low, with $\approx 10^{-4}$
giving a reasonable fit only to the observed high-energy cosmic ray data.
Using the compression ratio of $R = 3.2 \pm 0.2$ gives a proton spectrum of
$E^{-2.36 \pm 0.12}$ implying for the same flux at a few $10^{19}$ eV a
combined efficiency which is in the range $0.03 - 0.1$ (Fig. 5 in Kang et
al. 1997) and so consistent with estimates for supernova remnants {
(e.g. Drury et al. 1989; V\"olk 1997)\nocite{drury89,voelk97}}.

\subsection{\label{Sec:Eff}Shock Efficiency}

A rough estimate of the gas densities on both sides of the shock sphere can be
gained by extrapolating the electron density profile of the Coma cluster, viz.
$n_{\rm e} = n_{\rm e,o}\,[1+(r/r_{\rm core})^2]^{-3\beta/2}$, with the
parameters $n_{\rm e,o} = 3\cdot10^{-3}\,h_{50}^{-1/2}{\rm cm^{-3}}$, $r_{\rm
core}= 400\,h_{50}^{-1}\, {\rm kpc}$, and $\beta=0.75$ (Briel et
al.~1992)\nocite{briel92}.  This gives a gas density of the order $n_{\rm
e,2}\approx 10^{-5} \, h_{50}^{-1/2} {\rm cm^{-3}}$ at a cluster radius of
$4.8\,h_{50}^{-1}${ Mpc}. The density of the infalling matter should
therefore be of the order of $n_{\rm e,1} \approx 3.5\cdot
10^{-6}\,h_{50}^{-1/2}{\rm cm^{-3}}$. This is consistent with the somewhat
lower gas density in the intergalactic space of $\leq 10^{-6}\, {\rm cm^{-3}}$
far away from clusters, derived e.g. from radio observation of the giant radio
galaxy 1358+305 (Parma et al.~1996)\nocite{parma96}.

The projected extent of 1253+275 is $1.2\times0.4 \,h_{50}^{-2}\,{\rm Mpc^2}$
(Giovannini et al.~1991), but its real surface $S$ exposed to the accretion
flow might be larger.  The kinetic power of the infalling matter on the relic
\begin{equation}
\label{eq:Qflow}
Q_{\rm flow} = \frac{1}{2}\,n_{\rm e,1}\,m_{\rm p}\,\tilde{V}_{\rm s}^2 V_{\rm
s}\,S \approx 5\cdot 10^{43} \,h_{50}^{-3/2}\, {\rm erg\,s^{-1}}\,\,,
\end{equation}
is mainly transformed into thermal energy, but some fraction is
converted into relativistic particles. $\tilde{V}_{\rm s} = V_{\rm
s}\,(R-1)/R$ is the velocity of the infalling matter measured in the
cluster's inertial frame. The integrated radio power between 10 MHz
and 10 GHz of { $8\cdot 10^{40}\,h_{50}^{-2}\, {\rm erg\,s^{-1}}$
(Giovannini et al.~1991)\nocite{giovannini91}} is three orders of
magnitude lower, and can be easily powered by the dissipative
processes in the shock.  Assuming a field strength of $1 \mu$G implies
that only 10\% of the electron radiation losses are synchrotron
emission, the rest are inverse Compton losses by scattering of
microwave background photons. The amount of energy loss of the
electrons $Q_{\rm loss}\approx 8 \cdot 10^{41}\,h_{50}^{-2}\, {\rm
erg\,s^{-1}}$ and the power of the flow $Q_{\rm flow}$ onto the relic
gives the necessary minimal efficiency of shock acceleration. The
efficiency needs to be higher than $1\%$ in order to account for the
radiative energy requirements and for escaping electrons. This number
might be compared to the efficiency of shock acceleration in other
astrophysical objects: It is believed that supernova blast waves have
efficiencies of $1\%- 10\%$ for the acceleration of protons as is
necessary in order to explain the galactic cosmic rays below $10^{15}$
eV by supernovae { (e.g. Drury et al. 1989; V\"olk
1997)\nocite{drury89,voelk97}}. This shows that the assumed cluster
efficiency is reasonable.
\subsection{Electron Spectrum\label{sec:diffusion}}
We apply the theory of plane-parallel shock acceleration, because of the large
radius of the shock sphere. { We mark upstream quantities with index 1, and
downstream quantities with index 2.} We use a momentum independent diffusion
coefficient, because of the success of this simplification in other
circumstances (Biermann 1993, 1996; Wiebel-Sooth et al. 1997).
\nocite{biermann93} \nocite{biermann96} \nocite{wiebel97}
The distribution function of the accelerated electrons cooling by inverse
Compton and synchrotron emission is
\begin{equation}
f(x,p) = C \, p^{-q} \, \exp(-p/p^{*}(x))
\end{equation}
(Webb et al.~1984\nocite{webb84}), where the normalization $C$ can be
approximated by being independent of the position $x$, and $q=
3R/(R-1)$ is the spectral index of the injection electron distribution
in three-dimensional momentum space. The cutoff momentum is
\begin{eqnarray}
p^*(x) &=& \frac{4 \, p_{\rm o}}{q/a_{1} + (q-3)/a_{2} + 
|x|U/(\kappa\,a)}\nonumber\\
&=:& ( F + |x|G)^{-1}\,\,,
\end{eqnarray}
with $U$ denoting the flow velocity in the inertial frame of the shock
($U_1 =V_{\rm s}$ and $U_2 = U_1/R$), $\kappa$ is the diffusion
coefficient, $p_{\rm o}$ is the injection momentum into the
acceleration mechanism, and $a$ is the ratio between loss and
acceleration time scale:
\begin{equation}
a = \frac{U^2\,m_{\rm e} \,c\,\tau_{\rm loss}}{4\,\kappa\,p_{\rm o}}
\end{equation}
The energy loss time scale $\tau_{\rm loss}\,m_{\rm e}c/p$ of an electron with
momentum $p$ is dominated by synchrotron and inverse-Compton losses with:
\begin{equation}
\tau_{\rm loss}^{-1}= \frac{4 \sigma_{\rm T}}{3 m_{\rm e} c}
\left(\frac{B^2}{8\pi}+ \varepsilon_{\rm ph} \right)\,\,,
\end{equation}
where $\varepsilon_{\rm ph}$ is the energy density of the photon field, which
is dominated by the microwave background.  In Sect. \ref{depolarization} it is
argued that the main fraction of the synchrotron emitting volume belongs to
the post-shock region. Thus the electron population, integrated over the
synchrotron emitting relic volume ${\rm Vol} =SD$ (where $S$ is the surface
area, and $D$ is the thickness), can be assumed to be
\begin{eqnarray}
f(p) &=& S\,\int_0^D dx\,f(x,p)\\ &=& \frac{SC}{G_2}
     \,p^{-(q+1)}\,\exp(-pF)\,\left[ 1-\exp(-p G_2 D)
\right] \,\,. \nonumber
\end{eqnarray}
This spectrum has a break at $p_{\rm break} = 1/(G_2D)$ between the
spectral index $q$ and $q+1$, and a cutoff at $p_{\rm cut}=1/F$. 
The  break in the momentum spectrum at
\begin{equation}
\label{eq:break}
p_{\rm break} = \frac{m_{\rm e}\,c\,U_2\,\tau_{\rm loss,2}}{D} 
\end{equation}
leads to a break in the radio spectrum at $\nu_{\rm break} = 3 e B_2 \,p_{\rm
break}^2/(2 \pi m_{\rm e}^3 c^3)$. Using this relation and solving Eq.
\ref{eq:break} for the ratio $D/U_2$, which is the time the magnetized plasma
needed to pass the shock, and therefore the age of the relic as a tracer
of the shock structure, we get
\begin{equation}
\label{eq:tage}
t_{\rm age} = \frac{D}{U_2} = 
\sqrt{\frac{3 e B_{\rm 2,eq} m_{\rm e} c}{2 \pi \nu_{\rm break}}}
\left[ \frac{4}{3} \sigma_{\rm T} \left(\frac{B_{\rm 2,eq}^2}{8\pi}+
\varepsilon_{\rm ph} \right) \right]^{-1}.
\end{equation}
We use a thickness of $D= 100\, h_{50}\,{\rm kpc}$, roughly $\frac{1}{4}$ of
the width of the projected structure of 1253+275, due to compression.  The age
of 1253+275 being a shock-tracer estimated from the kinematics $t_{\rm
age,kin}= D\,R/V_{\rm s} = 1.5\cdot 10^{8} \,{\rm yr}\, h_{50}^{-1}$, and that
following from Eq. \ref{eq:tage}: $t_{\rm age, break} > 4\cdot 10^8 \,{\rm
yr}$, using the lowest observation frequency of 151 MHz as an upper limit to
the break frequency $\nu_{\rm break}$, do not fit exactly. { Regarding the
approximations of this theory, and the uncertainties of the observational
quantities} only an order of magnitude accordance can be expected.  But since
these two estimates of the relic age { depend on different observational
quantities} ($t_{\rm age, kin}(kT_{\rm obs}, \alpha , D)$ and $t_{\rm age,
break}(\nu_{\rm break}, B_{\rm 2,eq})$), this is a further test of the theory,
which could have failed by orders of magnitude. { Reasons for a discrepancy
might also be hidden in neglected properties of the acceleration mechanism. For
instance if the relic has passed the shock already, the diffusion coefficient
directly behind the shock $\kappa_{\rm 2, shock}$ differs from that within the
relic $\kappa_{\rm 2,relic}$, and the ratio $\kappa_{\rm 2,relic}/\kappa_{\rm
2, shock}$ would enter Eqs. \ref{eq:break} and \ref{eq:tage}.}

The ratio between break and cutoff momentum of 1253+275
\begin{equation}
\zeta = \frac{p_{\rm break}}{p_{\rm cut}} = \frac{F}{G_2D} =
\frac{\kappa_2}{U_2 D} \left( q \frac{a_2}{a_1} + q-3 \right)
\end{equation}
must be smaller than 0.18, since the radio spectrum does not exhibit any break
or cutoff { over a frequency range of about $1\frac{1}{2}$ orders of
magnitude}. The term in brackets is in the range $1- 3$ and can be
neglected for an order of magnitude estimate of the diffusion coefficient
behind the shock:
\begin{equation}
\label{eq:kappa2}
\kappa_2 \approx  V_{\rm s}\,D\,\zeta / R < 3 \cdot 10^{30} \,h_{50}\,{\rm
cm^2\,s^{-1}} \,\,.
\end{equation}

\section{\label{polarization}Field Compression and Radio Polarization}
\subsection{\label{Sec:general}General Considerations}

The high degree of polarization of relic sources should result from
the compression, which aligns unordered magnetic fields with the shock
plane. If the shock processed relic is seen at some angle
$\delta>0^\circ$ between the line-of-sight and the normal of the shock
front, the field structure projected onto the plane of sky shows a
preferential direction. Thus, the resulting radio polarization depends
on $\delta$ and $R$. The direction of the magnetic field should appear
to be perpendicular to the line connecting the relic and the cluster
center, because of this compression and projection effects. This is
the case for 1273+275 (Andernach et al.~1984)\nocite{andernach84}.

In the following we calculate the magnetic field compression and alignment,
and the resulting radio polarization of an unordered magnetized plasma blob in
an unmagnetized gas flow going through a shock.
The (complex) polarization of synchrotron radiation of an isotropic
distribution of electrons within a magnetic field seen from the
$z$-direction is given as (Burn 1966)\nocite{burn66}
\begin{eqnarray}
P &=& \frac{\gamma +1}{\gamma +\frac{7}{3}}
\,\frac{B_y^2-B_x^2-2iB_xB_y}{B_y^2+B_x^2}\,\,.
\end{eqnarray}
The spectral index of the electrons is $\gamma = 2\alpha +1$, where $\alpha$
is the spectral index of the radio emission. In order to average over regions
with different field orientations and strength, we have to take into account
the different emissivities
\begin{equation}
\varepsilon_{\rm sync}= a_{\rm sync}\, ({B_y^2+B_x^2})^{\frac{\alpha
+1}{2}}\,\,. 
\end{equation}
Burn has argued that the case $\alpha =1$ is a valid simplification, which is
justified especially in our case, where $\alpha\approx 1$. Averaging the
polarization (denoted by $<...>$) over a (point symmetric) distribution of
fields, weighted with the relative emissivity $\varepsilon_{\rm
sync}/<\varepsilon_{\rm sync}>$, gives the observed integral polarization:
\begin{eqnarray}
\label{eqpol}
<P> &=& \frac{\gamma +1}{\gamma +\frac{7}{3}}
\,\frac{<B_y^2>-<B_x^2>}{<B_y^2>+<B_x^2>}\,\,.
\end{eqnarray}
A magnetic flux tube with field strength $B_1$ oriented at some
angle $\theta_1$ to the normal of the shock front will be bent and amplified
by a compression $\tilde{R}$. From flux conservation it follows that
\begin{eqnarray}
\label{theta} \tan\theta_2&=& \tilde{R}\,\tan\theta_1\\
\label{B2}    
B_2&=& B_1\,\sqrt{\cos^2\theta_1 + \tilde{R}^2\,\sin^2\theta_1}\,\,.
\end{eqnarray} 
Observing this field, which is given by its spherical coordinates $(B_2,
\theta_2, \phi_2=\phi_1)$, at the viewing angle $\delta$ results
in
\begin{equation} \label{vecB}
\left( 
\begin{array}{c}
B_x\\B_y\\B_z
\end{array}
\right) = 
\left( 
\begin{array}{c}
\cos\delta \cos\phi_2 \sin\theta_2 - \sin\delta \cos\theta_2\\
\sin\phi_2 \sin\theta_2\\
\sin\delta \cos\phi_2 \sin\theta_2 + \cos\delta \cos\theta_2
\end{array}
\right)
B_2 \,\,.
\end{equation}
Eq. \ref{eqpol} can now be integrated, assuming an isotropic distribution of
the unshocked fields $\vec{B}_1$ and choosing the compression ratio of the
fields $\tilde{R}$, which depends on the shock compression ratio $R$ and the
field orientation. We discuss two complementary extreme cases, which give
similar polarizations.
\begin{figure}[t]
\psfig{figure=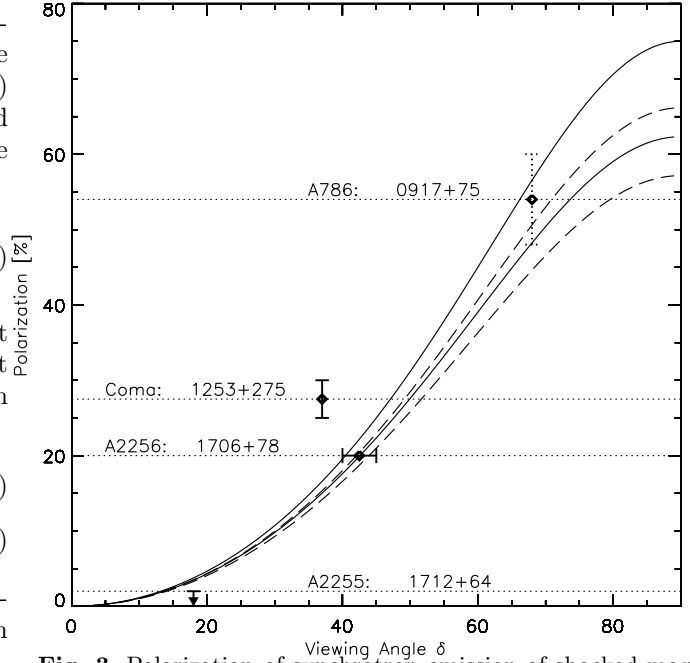,width=0.5\textwidth}
\caption[]{\label{pol:ps}Polarization of synchrotron emission of
shocked magnetic fields as a function of the viewing angle $\delta$.  The
strong field case is given by solid lines, the weak field case by dashed
lines. The upper pair of curves (solid \& dashed) corresponds to a radio
spectral index of $\alpha=1$, and therefore $\gamma=3$ and $R=4$. The lower
lines correspond to $\alpha=1.5$, $\gamma=4$ and $R=2.5$. Observed
polarizations of radio relics are plotted above the viewing angle predicted
from the accretion shock theory (1253+257, 1712+64), simulations of a cluster
merger (1706+78), or -- if no data is available -- at their best-fit position
(0917+75). The large projected radius of 0917+75 indicates a large viewing
angle, consistent with the best-fit position. Uncertainties in the angles are
large.}
\end{figure}
\subsection{Weak Fields\label{sec:weakB}}

If the magnetic pressure of the relic is small compared to the internal gas
pressure the compression of the magnetized regions is equal to the compression
of the accretion shock: $\tilde{R}=R$. The calculation of the integrals of
Eq. \ref{eqpol} is straightforward if the ensemble average is done over the
isotropic, unshocked fields, transformed via Eqs. \ref{theta}, \ref{B2}, and
\ref{vecB} to the shocked fields in the observer's coordinate system:
\begin{equation}
\label{eq:Pweak}
<P_{\rm weak}> = \frac{\gamma +1}{\gamma +\frac{7}{3}}
\,\,\, \frac{\sin^2\delta }{\frac{2\, R^2}{R^2-1}-\sin^2\delta}\,\,.
\end{equation}
Inserting numbers of the relic 1253+275 ($\delta=37^\circ$, $R= 3.2\pm 0.2$,
and $\gamma = 3.36\pm 0.122$, derived from $\alpha =1.18\pm 0.06$) gives a
polarization of $<P_{\rm weak}> = 15\%$. The observed polarization of $P_{\rm
obs} = 25\%- 30\%$ (Giovannini et al.~1991) is higher and might indicate
an intrinsically ordered field structure of the relic { as could be left
behind from ordered field structures within an unshocked radio lobe or tail as
a progenitor.  Or it is due to} a larger viewing angle of $\delta= 48^\circ
- 52^\circ$, corresponding to an accretion shock radius of $r_{\rm s}=
4.0- 3.6\,h_{50}^{-1}\,$Mpc instead of $r_{\rm s}=
4.8\,h_{50}^{-1}\,$Mpc, as given by Eq. \ref{eq:rs}. Regarding all the
simplifying assumptions made to derive both estimates (Eqs. \ref{eq:rs} and
\ref{eq:Pweak}) of the shock radius, especially the assumed spherical
symmetry, we feel that the derived values are quite consistent.

\subsection{Strong Fields\label{strongB}}

If the relic is supported by magnetic pressure only, then the compression of
flux tubes differs from the compression of the unmagnetized surrounding medium:
The progenitor of the cluster relic, an extended radio lobe, was in pressure
equilibrium with its surrounding gas in the upstream region. After the radio
plasma has passed the shock, it expands or contracts rapidly, depending on the
field orientation, in order to achieve pressure equilibrium with its
environment again, since the downstream flow (in the rest frame of the shock)
is subsonic. This relates the upstream and downstream fields by $B_2^2/B_1^2 =
P_2/P_1$. Eqs. \ref{P} and \ref{B2} then give the field compression factor
as a function of the field orientation:
\begin{equation}
\tilde{R}(R,\theta_1) = \sqrt{\frac{\frac{4R-1}{4-R} - \cos^2\theta_1 }{
\sin^2\theta_1}} 
\end{equation}
The polarization in the strong field case is then
\begin{equation}
\label{eq:Pstrong}
<P_{\rm strong}> = \frac{\gamma +1}{\gamma +\frac{7}{3}}
\,\,\, \frac{\sin^2\delta }{\frac{2}{15}\,\,\frac{13 R-
7}{R-1}-\sin^2\delta} \,\,.
\end{equation}
Inserting the numbers of 1253+275 gives { $<P_{\rm strong}> =
16\%$.}  This is similar to the weak field case, also demonstrated by
Fig. \ref{pol:ps}.  Magnetic fields of $1.9 \, h_{50}^{-1/4} \, \mu$G
are necessary within 1253+275 in order to support this structure
against the thermal pressure at a cluster radius of $4.8 \,
h_{50}^{-1}\, $Mpc, assuming a density and temperature as given in
Sect. \ref{compr} and a magnetic pressure of $B^2/(8\pi)$. This rough
estimate depends on an extrapolation of the central density profile of
the cluster, and has therefore a large error. The magnetic field
strength could be lower if nonmagnetic pressure such as thermal gas
and/or relativistic particles also support the radio structure. This
number is higher than the magnetic field strength derived from minimum
energy arguments of the synchrotron emitting plasma of
$0.5\,h_{50}^{2/7}\,\mu$G measured by Giovannini et
al.~(1991)\nocite{giovannini91}. Since these authors use a source
depth of $400\,h_{50}^{-1}\,$kpc along the line-of-sight, while we
expect only $125\,h_{50}^{-1}\,$kpc due to the compression of the
source by the shock, we can correct this number for our model to
$0.7\, h_{50}^{2/7}\,\mu$G. Using the rise of the thermal and (here in
the strong field case) of the magnetic pressure given by Eq. \ref{P}
we find that the field strength of the unshocked relic plasma needs to
be { $0.18 \,\mu$G}, if the equipartition value gives the relic
field strength, or $0.5 \,\mu$G if only magnetic pressure has to
support the relic. Field strengths of the pre- and after-shock region
should be within these ranges.

\subsection{Depolarization\label{depolarization}}

A somewhat stronger difference between the polarization in the strong and weak
field case might arise from { internal} Faraday depolarization, which
should only appear in the weak field case, since the relic has to be filled
with a sufficient amount of thermal gas. { Internal depolarization} should
be strongest for large viewing angle $\delta$, due to the larger number of
field lines aligned with the line-of-sight.

We believe that the strong field case is a better description of the remnant
of a radio galaxy's lobe, since we find rough equivalence between the measured
equipartition field strength and predicted magnetic fields from pressure
equilibrium with the surrounding gas. Therefore, { internal} depolarization
should be weak. We do not attempt any estimate because of the large
uncertainties entering the calculation such as coherence scale of the fields,
the true field strength, and gas density in the relic.

{ External Faraday depolarization can lower the observed polarization if the
relic is seen through the central magnetized regions of the cluster. The degree
of depolarization would depend on the field strength and reversal scale within
the intra-cluster medium, where both quantities have large uncertainties (Kim
et al.~1986\nocite{kim86}; Feretti et al.~1995\nocite{feretti95}; En{\ss}lin et
al.~1997\nocite{ensslin97}; En{\ss}lin \& Biermann 1997\nocite{ensslin97a}).
In order for external depolarization to occur the relic has to be on the back
side of the cluster, and its viewing angle $\delta$ must be low. Since then the
polarization of the relic would be very small anyway, we do not need to correct
for this.}

We note that, since the observed polarization of 1275+273 is high, nearly all
of the synchrotron emission has to come from the ordered fields of the
post-shock region { as assumed in Sect. \ref{sec:diffusion}}, otherwise the
polarization would be lower.

\section{X-Ray Emission from Inverse Compton Scattering}

The same electrons that emit the synchrotron radiation of the relic should
inverse-Compton scatter microwave background photons to the X-ray
range. Goldschmidt \& Rephaeli (1994)\nocite{goldshmidt94} calculated the flux
from 1253+275 to be:
\begin{equation}
F_{\rm X}(\varepsilon_\gamma)= \frac{7.1\cdot 10^{-5}}{\rm
cm^{2}\,s\,keV} \left( \frac{\varepsilon_\gamma}{\rm keV} \right)^{-2.18}
\left( \frac{B_2}{\mu\rm G} \right)^{-2.18} \,\,.
\end{equation}
Since only a very low thermal X-ray background is produced at such a large
cluster radius, this inverse-Compton flux should be best detectable at keV
energies. The predicted flux in the ROSAT 0.5-2.4 keV band is $F_{\rm
X,0.5-2.4keV} = 1.2\cdot 10^{-4}\,(B_2/{\rm \mu G})^{-2.18}\, {\rm
cm^{-2}\,s^{-1}}$. The observed X-ray energy flux from the Coma cluster is
plotted in Fig. 2 of { White et al.  (1993)\nocite{white93}}, and is $9
\cdot 10^{-12}\,{\rm erg\, cm^{-2}\,s^{-1}\,deg^{-2}}$ at the position of the
relic (assuming a emission temperature of 8 keV). This can be translated into
an observed flux of $F_{\rm X,0.5-2.4keV}= 4\cdot 10^{-4}\,{\rm
cm^{-2}\,s^{-1}}$ coming from the field covered by the relic. Comparing this
number with the predicted flux implies a field strength of $B_2 > 0.6\,\mu$G,
since most of the observed X-ray emission should belong to the X-ray blob
centered on NGC 4839. This field strength is in good agreement with the field
strength derived by minimum energy arguments and pressure equilibrium, given
in Sect. \ref{strongB}. It has to be noted though that if a break in the
synchrotron spectrum below the observed range (150 MHz -- 4.75 GHz) indicates
a break in the relativistic electron population, the resulting inverse-Compton
flux in the ROSAT band would be largely reduced, since radio frequency $\nu$
translates into inverse Compton energy via $\varepsilon_{\gamma}= 1\,\,{\rm
keV} (B_2/{\mu
\rm G})^{-1}\,(\nu/{10\rm \,MHz})$ in the monochromatic approximation of the
synchrotron and inverse Compton formulae (Blumenthal \& Gould
1970).\nocite{blumenthal70} Assuming a break in the radio spectral index of 0.5
at 10 MHz or 100 MHz still gives lower limits of $B_2>0.4\,\mu$G or
$>0.2\,\mu$G, respectively. The latter value can be regarded as a hard lower
limit to the relic field strength.

1253+275 should be an interesting object for future X-ray telescopes, such as
XMM and AXAF, because its large-scale inverse Compton flux might be detected,
or stronger limits to the magnetic fields and/or the low-energy electron
spectrum can be derived. The radio halo of the Coma cluster should have a
larger inverse-Compton flux, but there the keV range is polluted by thermal
emission and has to be avoided by using other energy bands (Rephaeli et
al.~1994,\nocite{rephaeli94} En{\ss}lin \& Biermann 1997)\nocite{ensslin97a}.

\section{\label{sec:origin}The Origin of Relics}
\subsection{Radio Galaxies\label{sec:radio_galaxies}}
Relics are believed to be remnants of former radio galaxies. The problem that
the energy loss time scale is frequently shorter than the time elapsed since
the last resupply with energetic electrons from a progenitor radio galaxy is
solved in our approach due to the acceleration by the large-scale accretion
shocks. Therefore the progenitor radio galaxy needs not to be within the
distance from the relic a galaxy can travel with a typical cluster velocity
within a cooling time of the electrons. In order to explain the polarization
by field compression, it is necessary though that the radio plasma from which
the present day radio emission results was injected within the upstream
region. The large spatial extent of relics might result from the much lower
pressure in the upstream regions, so that the radio lobes could expand over a
larger volume than a similar radio galaxy would have occupied in the
intra-cluster medium inside the accretion shock.

It is interesting to remember that the minimum magnetic pressure of 1253+275 is
near equipartition with the expected thermal pressure at the shock radius
(estimated using an extrapolation of the central density profile), whereas
minimum pressures of radio lobes of active radio galaxies in the interior of
galaxy clusters usually are much lower than the gas pressure at their radii,
even if projection effects are taken into account (Feretti et
al.~1992)\nocite{feretti92}. Several effects can explain this latter pressure
imbalance, despite the possibility that minimum energies do not describe the
physical conditions:
\begin{itemize}
\item
a filling factor of the observed radio plasma of a few
percent 
\item
the ubiquitous presence of relativistic protons within the lobes,{
dominating the energy density of relativistic particles}
\item
the existence of a significant number of low-energy electrons radiating below
the adopted low-frequency cutoff of the minimum energy analysis
\item
a significant amount of thermal gas within the lobes.
\end{itemize}
Under the hypothesis that relics are old { lobes}, the filling factor and
also the low-energy electron population should be similar in both types of
sources, and therefore not responsible for differences of cluster relics and
lobes in the central regions of clusters. Thermal gas remains to be the
explanation, and also relativistic protons, because the possible rich initial
proton population of the lobes might be escaped due to the larger age and the
higher cosmic ray diffusion coefficient of relics compared to lobes inside
clusters.

Giovannini et al.~(1985)\nocite{giovannini85} propose the Coma cluster galaxy
IC 3900 to be the origin of 1253+275, which should have moved to another
position on the plane of the sky. We point out that the narrow-angle tail (NAT)
radio galaxy NGC 4789 is a very attractive possible origin, too. First, a
bridge of radio emission connects it with the central ridge of the relic
(Giovannini et al.~1991\nocite{giovannini91}), indicating a geometrical
connection. Second, 1253+275 lies on the (projected) line between NGC 4789 and
the center of the Coma cluster. NGC 4789 is therefore an ideal candidate for
being the source of magnetized plasma injected upstream into the accretion
flow. The NAT structure of NGC 4789 might result from the ram-pressure of the
infalling matter on the radio jets, enhanced by an ascending movement of the
galaxy. Relativistic electrons from the jets should be convected by the flow
within the radio bridge to the accretion shock, where they are reaccelerated
and radiate. The line-of-sight velocity of NGC 4789 is higher by $1179\,{\rm
km\,s^{-1}}$ than Coma's Hubble velocity (Venturi et
al.~1988)\nocite{venturi88}. This implies, if NGC 4789 is really located
upstream above 1253+275, that we see a relic on the back side of the accretion
shock sphere. This is sketched in Fig.~\ref{skizze.ps}. { If NGC 4789 is on
a radial ascending orbit, its velocity should be 1480 km/s (assuming $\delta
=37^\circ$), similar to that of the infalling matter measured in the clusters
rest frame $\tilde{V}_{\rm s}= 1400$ km/s. The radio plasma, which is reaching
the shock today, was thus injected into the stream half a Gyr ago, given that
its movement is only due to convection by the accretion flow.}

\subsection{Large-Scale Fields}
A very different nature { of the origin of the magnetic fields in the
relic} is imaginable: from simulations of magnetic field generation and
amplification by turbulence in the flows of large-scale structure formation
field strengths of the $0.1\,\mu$G level within cosmological filaments are
predicted by Biermann et al.~(1997). The predicted field strengths at cluster
accretion shocks are $B_2 \approx
\mu$G, being in energetic equipartition with the thermal energy
density. In the inner regions of clusters, field strengths at the
$10\,\mu$G level are expected, as En{\ss}lin et
al.~(1997)\nocite{ensslin97} argue to be injected from radio galaxies,
presuming that their jets also contain energetic protons.  If field
values of $B_2 \approx \mu$G are typical for the accretion shock
region, it has to be explained what is special at the positions of
relic sources. In the case of 1273+275 this should be the presence of
NGC 4789 providing preaccelerated electrons injected into the
accretion flow, efficiently reaccelerated at the shock front below the
galaxy.

It is remarkable that the 1253+275 complex is in the elongation of the Coma
cluster, which itself appears to align with Coma/Abell 1367 filament of
galaxies as described by Fontanelli (1984)\nocite{fontanelli84}.  It is
possible that the position of the relic traces the working surface of a
large-scale stream flowing out of this filament and hitting the sea of cluster
gas. The observed radio bridge connecting 1253+275 and the radio halo Coma C
(Kim et al.  1989)\nocite{kim89} then might be a deposition of magnetic fields
from this stream, illuminated by relativistic electrons accelerated in the
shock.

\section{Other Relics\label{Sec:other}}
\subsection{General Remarks}

A handful of other cluster relics are known. Their properties and that of
their host clusters are collected in the upper part of
Tab. \ref{tab:relics}. According to the theory discussed in this paper other
quantities are calculated and given in the lower part of
Tab. \ref{tab:relics}.  { Average values for 1253+275 are also given in
Tab. \ref{tab:relics}, but the very recently discovered relic 2010-57
(R\"ottgering et al 1997)\nocite{roettgering97} is only discussed in the
text.}

For four relics, namely 0917+75, 1253+27, 1712+64, and 1706+78, polarization
measurements are available. In three cases these polarizations fit well into
the prediction of the accretion shock theory (see Fig. \ref{pol:ps}). In the
case of 1706+78 in A2256 the observed polarization is much higher than
expected. This can be understood if the shock is due to an on-going merger
event (Fabian \& Daines 1991)\nocite{fabian91}, and therefore located at a
smaller cluster radius. The accretion shock theory predicts a polarization of
{ 28\% for the relic 2006-56, 10\% for 2010-57, 4\% for 1140+203 and
1401-33}, and 1\% for 0038-096.

The predicted shock velocity $V_{\rm s,predicted}$ is in all accretion shock
examples lower than that of the accretion shock theory of Kang et al. (1997)
by { 16\%} on average. This could be due to an assumed temperature drop of
a factor of less than two from the cluster center to the accretion shock
radius, or due to the need of some recalibration of the accretion shock {
parameters}.

Looking at the distribution of the predicted viewing angle $\delta$ one
recognizes that large angles are underrepresented. In an isotropic distribution
half of the relics should be seen at a viewing angle above 60$^\circ$. This is
clearly a selection effect, since the sensitive radio surveys of clusters with
good angular resolution which are needed for the detection of these sources,
usually do not reach projected radii of 5 Mpc, as would be necessary in order
to include the whole accretion shock.  We expect therefore that undiscovered
relics, with strong radio polarizations, are waiting in these outer regions of
clusters for their exploration.


\ifthenelse{
\value{page}= 2 \or 
\value{page}= 4 \or
\value{page}= 6 \or
\value{page}= 8 \or
\value{page}= 10 \or
\value{page}= 12 \or
\value{page}= 14 \or
\value{page}= 16 \or
\value{page}= 18
}{\begin{figure*}[tbp] \end{figure*} \setcounter{dbltopnumber}{10}
\setlength{\dbltextfloatsep}{1.5em} 
}{ }

\begin{table*}[tbp]
\begin{tabular}{|l|l|l||c|c|c|c|c|c|c|c|}  \hline 
\input{relics.tab}
\end{tabular}
\end{table*}
\begin{table*}
\caption[]{\label{tab:relics}The upper part of the table is a
collection of measurements of properties of cluster radio relics and their
host clusters. The references are indicated in the third column in the sequence
of the eight examples.  The lower part gives estimates, using the above
properties and the formulae of this paper, as indicated in the third column. A
discussion of the individual sources and various assumptions made in order to
get the given values can be found in Sect. \ref{Sec:other}.\\[0.5em]}
\end{table*}
\begin{table*}
\small
{\bf Observed Quantities:} 
$z$: redshift of the cluster.
$kT_{\rm obs}$: observed central gas temperature.
$n_{\rm e,o}$, $r_{\rm core}$, $\beta$: parameters of the $\beta$-model, 
describing the radial electron density profile of the cluster $n_{\rm e} =
n_{\rm e,o}\,[1+(r/r_{\rm core})^2]^{-3\beta /2}$.
$r_{\rm projected}$: radius of the position of relic, projected into the plane
of sky.
$\alpha$: spectral index of relic's radio emission.
$P$: average observed polarization.
$B_{\rm 2,eq}$: equipartition field strength of the relic, the index 2 refers
to the post-shock region, where to the visible part of the relic belongs. The
values of the literature are changed due to the effect of a smaller volume
than assumed there according to $B_{\rm 2,eq} \sim R^{2/7}$, where $R$ is the
compression ratio.
$D$: thickness, roughly estimated by dividing the observed projected thickness
by the compression $R$.
$S$: surface area of the relic. Since in all except one cases the relic
structure can not be deprojected, we use the observed projected surface as a
rough estimate.
$\nu_{\rm break}$: break frequency of relic's radio spectrum.
$\zeta$: ratio of break- to cutoff momentum of the electron population,
estimated from the ratio of break- and cutoff frequency of the radio spectrum
by taking the square root.
$Q_{\rm radio}$: relic's radio power, usually assuming a single power law
between 10 MHz and 10 GHz.\\[0.5em]
\end{table*}
\begin{table*}
\small
{\bf Estimated Quantities:}
$r_{\rm s}$: shock radius. $V_{\rm s}$: shock velocity. 
$\delta$: viewing angle given by the estimate of the shock
radius $r_{\rm s}$ and the projected radius of the relic $r_{\rm proj}$.
$R$: shock compression ratio, estimated from the spectral index.
$P_2/P_1$, $T_2/T_1$: pressure and temperature jump across the shock.
$n_{\rm e,1}$: upstream side electron density, derived from the
$\beta$-profile and the compression ratio.
$kT_1$: temperature of the infalling gas, derived from the post-shock
temperature $kT_2$, which is assumed to be half of the central temperature
$kT_{\rm obs}$ (see Sect. \ref{compr}). 
$V_{\rm s,predicted}$: shock velocity, predicted from the post-shock
temperature and the compression ratio.
$B_{\rm 1, predicted}$, $B_{\rm 2, predicted}$: up- and downstream field
strength which would be in pressure equilibrium with the surrounding gas.
$B_{\rm 1,eq}$: strength of the magnetic field of the progenitor
of the relic source, if the magnetic pressure was increased by
$P_2/P_1$ to $B_{\rm 2,eq}$ in the shock.
$Q_{\rm flow}$: kinetic power of the flow onto the surface of the relic.
$\eta_{\rm radio}$: radio efficiency $ Q_{\rm radio}/Q_{\rm flow}$.
$t_{\rm age,kin}$: age of the relic being behind the shock from $D/U_2$,
where $U_2= V_{\rm s}/R$ is the post-shock velocity.
$t_{\rm age, break}$: age of the relic being behind the shock from $D/U_2$,
where this ratio is derived from the radio break frequency.
$\kappa_2$: diffusion coefficient of the post-shock region.
$P_{\rm strong}$: polarization expected in the strong field case, for a
viewing angle $\delta$.
$\delta_{\rm strong}$, $r_{\rm s,strong}$: viewing angle and the shock radius
{ computed} to be consistent with the observed polarization $P$.\\[0.5em]
\newcounter{remref}
{\bf Remarks:}

\refstepcounter{remref}
$^{(\arabic{remref}\label{rem:A85})}$ The lower limit to the spectral index
translates into limits of dependent quantities. But the measured steep
spectral index might also be due to the appearance of the high frequency
cutoff, so that these limits become estimates.
\refstepcounter{remref}
$^{(\arabic{remref}\label{rem:s})}$ A spectral index smaller than unity
indicates, that the range observed in the radio is below the break
frequency. The spectral index above the break is steeper by 0.5.  A {
after-break} spectral index of $\alpha= 1.3$ was used in these formulae.  No
lower limit to the ratio $\zeta$ between cut and break frequency and therefore
no diffusion coefficient can be estimated in this case.
\refstepcounter{remref}
$^{(\arabic{remref}\label{rem:merge})}$ The results of the accretion
shock theory are $r_{\rm s}= 4.9 \,{\rm Mpc}\,h_{50}^{-1}$, $V_{\rm
s}= 2280\,$km/s, and $P_{\rm strong}= 0.78\%$. The latter is excluded by
the observed polarization of $P=20\%$. The accretion shock model
therefore fails; but as discussed in Sect. \ref{sec:A2256} the shock
should result from a merger. We therefore inserted the best-fit
numbers from a hydrodynamic simulation of such an event. $r_{\rm s}$
is set to $r_{\rm s,strong}$, $V_{\rm s}$ and $\delta$ are taken from
Roettiger et al.~(1995)\nocite{roettiger95}.
\refstepcounter{remref}
$^{(\arabic{remref}\label{rem:kTEA2256})}$ 
\refstepcounter{remref}
Since this shock is in the interior of the cluster, $kT_2 =kT_{\rm
obs}$ was used.
$^{(\arabic{remref}\label{rem:deproj})}$
For this relic a deprojection of the geometry was possible as described in the
text, and we therefore use these values.
\refstepcounter{remref}
$^{(\arabic{remref}\label{rem:kT})}$ The velocity dispersion given in
Fadda et al.~(1996) is translated into a temperature with the help of
the dispersion-temperature relation given in Bird et
al.~(1995)\nocite{bird95}.\\[0.5em]
{\bf References:  }
\cite{briel91}: Briel et al.~(1991),
\cite{briel92}: Briel et al.~(1992),
\cite{burns95}: Burns et al.~(1995), 
\cite{david93}: David et al.~(1993),
\cite{david95}: David et al.~(1995),
\cite{fadda96}: Fadda et al.~(1996), 
\cite{feretti96}: Feretti \& Giovannini~(1996),
\cite{feretti97}: Feretti et al.~(1997),
\cite{gavazzi78}: Gavazzi (1978),
\cite{gavazzi83}: Gavazzi \& Trinchieri (1983),
\cite{gavazzi87}: Gavazzi \& Jaffe (1987),
\cite{giovannini91}: Giovannini et al.~(1991),
\cite{goss82}: Goss et al.~(1982), 
\cite{goss87}: Goss et al.~(1987),
\cite{harris93}: Harris et al.~(1993), 
\cite{jones84}: Jones \& Forman (1984), 
\cite{joshi86}: Joshi et al.~(1986),
\cite{knopp96}: Knopp et al.~(1996),
\cite{lima97}: Lima Neto et al.~(1997),
\cite{roettgering94}: R\"{o}ttgering et al.~(1994), 
\cite{roettgering97}: R\"{o}ttgering et al.~(1997), 
\cite{zabludoff93}: Zabludoff (1993)
\vspace{2em}
\end{table*}

\setlength{\dbltextfloatsep}{0em} 

\subsection{Abell 85: 0038-096}
The spectral index of 0038-096 is not settled in the literature: Slee
\& Siegman (1983)\nocite{slee83} give $\alpha= 1.60$, Reynolds (1986)
gives $\alpha=2.90$, Slee et al.~(1994)\nocite{slee94} 2.37, Joshi et
al.~(1986)\nocite{joshi86} 3.03 { above} a break at 300 MHz,
whereas Feretti \& Giovannini (1996)\nocite{feretti96} state that
$\alpha> 1.5$. We base our estimates mainly on the numbers of the
latter authors. It is possible that the observed strong steepening is
in fact a smeared-out cutoff, resulting from the limited acceleration
power of a weak accretion shock.  Since the projected relic position
is close to the cluster center the viewing angle is close to zero and
therefore no visible polarization resulting from field compression is
expected. The necessary radio efficiency $\eta_{\rm radio}$ of the
shock is $>1\%$ and thus reasonable.  The predicted magnetic field
strength $B_{\rm 2,predicted}= 2.6 \,\mu$G$\,h_{50}^{-1/4}$ from
assuming pressure equilibrium between the post-shock gas and the relic
is again higher than the equipartition (or minimum energy) value given
by Feretti \& Giovannini (1996)\nocite{feretti96} $B_{\rm 2,eq} <
1.4\,\mu${ G}$\, h_{50}^{2/7}$ (corrected by $R^{2/7}$).  Lima Neto
et al. (1997)\nocite{lima97} report the appearance of an X-ray blob in
the X-ray image of A85 centered on 0038-096, possibly inverse-Compton
scattered microwave background photons. If so, the magnetic field
strength could be derived directly from an estimate of this X-ray
excess.

\subsection{Abell 786: 0917+75}

The megaparsec cluster radio relic 0917+75 is located 5
Mpc$\,h_{50}^{-1}$ away from the center of A786 within the plane of
sky. Unfortunately no temperature or velocity dispersion of A786 is
available in the literature. Thus, the expected shock radius and
therefore the polarization cannot be predicted. However the very
peripheral position indicates a large viewing angle, consistent with
the high degree of the observed polarization of $48 - 60\%$. A
viewing angle of $72^\circ - 63^\circ$ corresponding to an
accretion shock radius of $5.6 - 5.3\,$Mpc$\,h_{50}^{-1}$ would
be in agreement with this polarization. Such a large shock radius
indicates a deep gravitational potential. The corresponding virial
temperature $13 - 15$ keV (Eq.
\ref{eq:rs}) is in the upper range of observed temperatures of clusters,
and therefore possible. However the shock structure might in fact be also due
to an accretion flow, which is centered on and stops at the boundary of the
supercluster Rood \#27, which A786 is a member of. This could explain the large
projected shock radius.

The direction of the projected magnetic field observed by radio
polarization is mainly perpendicular to the axis connecting the center
of A786 and 0917+75, as it should be if the field is compressed at a
spherical shock centered on the cluster. But a weakly visible spiral
structure in the projected fields could result from an intrinsic order
which the magnetic fields had before they passed the shock.

The relic 0917+75 exhibits a break in its spectrum (Harris et al.~1993). Below
300 MHz the spectral index is 0.6 and above it is steeper by the canonical
value 0.5, as is expected from a steepening of the spectral index of the
electron population by 1, predicted in Sect. \ref{sec:diffusion}. The
corresponding compression ratio of $R=3.5$ is close to the maximum possible
value $R=4$ of a strong nonrelativistic shock.

The projected peripheral position of 0917+75 allows a sensitive search for
inverse-Compton scattered microwave background photons in the X-ray
range. Harris et al.~(1995) observed the relic with the PSPC instrument on the
ROSAT satellite and got an upper limit to the flux in the 0.5-2.0 keV band of
$2.3\cdot10^{-14}\,{\rm erg \, cm^{-2}\,s^{-1}}$. This implies that the
magnetic field strength must be $B_{\rm 2} \geq 0.8 \,\mu$G, slightly larger
than $B_{\rm 2,eq}= 0.5 \,\mu$G$\, h_{50}^{2/7}$ derived from minimum energy
considerations and various assumptions, which were discussed in Harris et
al.~(1995)\nocite{harris95} and Sect. \ref{sec:radio_galaxies} of this
article.

\subsection{Abell 1367: 1140+203}	

The diffuse radio source found in A1367 is usually called a halo, but it is
more probably a cluster relic because of its noncentral location and
asymmetric shape. { The distance to the center of the galaxy distribution
is $0.8\, {\rm Mpc}\,h_{50}^{-1}$ (Gavazzi 1978)\nocite{gavazzi78}, but to
that of the X-ray emission it is $1.1\, {\rm Mpc}\,h_{50}^{-1}$ (see map in
Gavazzi et al. 1995)\nocite{gavazzi95}. We use the latter distance, since the
X-ray emission should be a better tracer of the gravitational potential. The
X-ray emission is elongated into the direction of the relic, possibly tracing
the influence of the large-scale structure potential, or the main direction of
the accretion flow onto the cluster, or both. The radio relic could be
contaminated or related to three close irregular galaxies, which have radio
trails behind them (Gavazzi \& Jaffe 1987)\nocite{gavazzi87}, indicating that
they are falling inwards. Strong on-going star formation, a recent supernova
in one of them, and young and abundant H II regions, which are aligned as if
they were formed by bow shocks (Gavazzi et al.~1995\nocite{gavazzi95}), could
be triggered by the sudden change of the environment of these galaxies during
the passage of the accretion shock. A starburst might increase the accretion
within these galaxies onto a possible central black hole (similar to the
argumentation of Wang \& Biermann 1998)\nocite{wang98} and therefore can
trigger the ejection of radio plasma.}

The measurements of the spectral index and radio flux of the relic differ
in the literature (compare Gavazzi 1987, Hanisch 1980, and Gavazzi \&
Trinchieri 1983) and therefore any derived numbers have to be used with
care.  Especially the radio power $Q_{\rm radio}$ is an extrapolation of a
very steep spectrum ($\alpha = 1.9$) down to 10 MHz and probably an
overestimate.  Due to the low expected viewing angle only a small
polarization of { 4\%} is expected. Gavazzi \& Trinchieri (1983)
detected some X-ray emission from the region of the relic.  Assuming this
to result from inverse Compton emission they estimate a magnetic field
strength of $1.2\,\mu$G, of the same order as the equipartition value of
$B_{\rm 2,eq}= 2\,\mu $G$ \,h_{50}^{2/7}$ and the predicted strength of
$B_{\rm 2,predicted} = 2.4\,\mu$G$\,h_{50}^{1/4}$ for pressure equilibrium
with the surrounding.

\subsection{Abell 2255: 1712+64}
The cluster A2255 has a number of similarities to the Coma cluster: there is
evidence that the cluster is in a state of merging (Burns et
al.~1995)\nocite{burns95}, a radio halo is present, and also a peripheral
radio relic without any obvious parent galaxy (Feretti et
al.~1997)\nocite{feretti97}.  The spectral index of the relic is fairly
constant along the structure. The upper limits to the polarization of 9\% at
20 cm and 2\% at 90 cm do not contradict the predicted polarization of 3\%
seriously, taking into account e.g. the rough assumptions which entered the
estimate of the shock radius, { and the possibility that external Faraday
depolarization has lowered the degree of the radio polarization if the relic
is seen through the denser magnetized central intra-cluster medium}. This
relic therefore fits also well into the accretion shock theory.

\subsection{Abell 2256: 1706+78\label{sec:A2256}}
The radio emission reveals a very complex situation within A2256: { a
mini-halo is visible in the map of Bridle \& Fomalont
(1976)\nocite{bridle76}}, and several head-tail radio galaxies can be
found. One NAT source (in the notation of { Bridle \& Fomalont
(1976)\nocite{bridle76}}: source C) is one of the longest head-tail sources
known. The tail extends over 700 kpc$\,h_{50}^{-1}$, is slightly bent and
exhibits a kink 570 kpc$\,h_{50}^{-1}$ behind the head { (R\"{o}ttgering
1994)\nocite{roettgering94}}. The bending can be understood as the
gravitational influence of the cluster mass to the trajectory of a fast radio
moving radio galaxy, at which the tail traces the path.  The kink might result
from a sudden change of the surrounding density or velocity field of the
background gas (R\"{o}ttgering et al.~1994), as it might happen if the radio
galaxy passes a shock. The projected position of the kink is within a very
extended irregular and sharp-edged region of diffuse radio emission in the
north-west region of the cluster. This region consists of two extended sources
(G and H in the notation of { Bridle \& Fomalont (1976)\nocite{bridle76}}),
which might be two independent relic sources with similar properties.

The spectral index of 0.8 of these possible cluster relics indicates that the
injection electron spectrum below the break in the momentum distribution are
observed. The spectral index of the synchrotron emission above the break
should be $\alpha= 1.3$. The break momentum (Eq. \ref{eq:break}) is highest
for a large shock velocity $V_{\rm s}$ and a thin magnetized post-shock region
(small $D$). The latter could explain why the emission region is relatively
sharp-edged. If the thickness $D$ is much smaller than we assumed, the
kinematic age of the relic $t_{\rm age,kin}$ would be much smaller than $10^8$
yr, in accordance with $t_{\rm age,break}< 1.4\cdot 10^{8}$ yr.  The
polarization of 20\% is much too high for the expected viewing angle
$\delta=9^\circ$ of a relic located at the canonical radius of the accretion
shock $r_{\rm s}= 5\,{\rm Mpc}\,h_{50}^{-1}$, which predicts only 1\%
polarization. A consistent viewing angle of $\delta_{\rm strong}=43^\circ$
would imply that if the normal of the shock front is pointing to the cluster
center, the shock radius is $r_{\rm s}= 1\,{\rm Mpc}\,h_{50}^{-1}$. This would
be in agreement with the average direction of the polarization, which is
oriented roughly perpendicular to this assumed normal of the shock front.  A
shock in the interior of the cluster can only be explained by a merger
event. Independent evidence for such a collision can be found in the X-ray
morphology, in the complex temperature structure and in the elongated galaxy
distribution. An extensive discussion of the observational signs reported in
the literature is given by Roettiger et al.~(1995). They compare these
observations with simulations of cluster mergers and conclude that the entire
data are best reproduced by a merger of two clusters with mass ratio 1:2 and a
relative velocity of 3000 km/s. The separation of the cluster centers is 0.75
Mpc$\,h_{50}^{-1}$. The smaller cluster is infalling from the north-west and
is approaching the observer. The angle between the line-of-sight and the
merger axis is in the best-fit model in the range of 40$^\circ$ to 45$^\circ$,
in agreement with $\delta_{\rm strong}= 43^\circ$ estimated from polarization,
{ without taking any possible influence of external Faraday depolarization
into account}.

The temperature $kT_{\rm obs}$ given in Tab. \ref{tab:relics} is an average
value of a complex distribution, biased by the heat production of the
shock. Since the infalling subcluster has a much higher temperature ($kT_{1}
\approx 4$ keV; $kT_2 =kT_{\rm obs}$ is assumed here) than the accretion
streams hitting all other known cluster relics ($kT_{1} \approx 1$ keV), the
moderate shock compression ratio $R=2.9$ at this shock, which has the highest
shock velocity of our sample, can be understood to result from the high
pressure of the upstream matter.

The predicted shock velocity derived from the up-stream temperature and
pressure jump is $V_{\rm s,predicted}= 3050\, {\rm km/s}$, being in agreement,
and completely independent of the above mentioned { merger} velocity from
the hydrodynamic simulations. { The merger velocity corresponds to the
infall velocity measured in the cluster rest frame $\tilde{V}_{\rm s}$, which
is usually lower by $(R-1)/R$ than the shock velocity $V_{\rm s}$. But in case
of a massive merger it is unrealistic to believe that A2256 behaves like a
single body, the downstream flow does surely not rest in the cluster reference
frame.  We therefore use a realistic velocity of $V_{\rm s}= \tilde{V}_{\rm s}
= 3000$ km/s.}

The line-of-sight velocities of the head-tail radio galaxies also point towards
the observer, and especially the obviously fast moving source C seems to
originate from the infalling cluster. R\"{o}ttgering et al.~(1994) have fitted
trajectories, calculated for a gravitational potential derived from X-ray data
to the tail of source C, and concluded that the initial velocity had to be in
the range of 2000-3500 km/s, in accordance with the value of the relative
cluster velocity given by Roettiger et al.~(1995). From the undisturbed
straightness of the source R\"{o}ttgering et al.~(1994) derive that the
turbulent motion of the cluster medium has a velocity $< 200\,$km/s, much too
low to explain the diffuse emission resulting from large-scale turbulent
motion. However, this is not in contradiction to the merger shock model for
this relic, where violent velocity changes occur only in the thin shock region,
which might have distorted the tail of source C at the position of the
kink. { We note, that a brightening of the radio tail is visible there, as
expected to result from the action of the shock onto the tail (R\"{o}ttgering
et al.~1994).  Source C should have moved from the kink within $2- 3\cdot
10^8$ years to its present day position. This is comparable or higher than a
typical cooling time of radio emitting electron, and explains the observed
fading of the tail.}

The details of the complex scenario might differ from the above picture in that
the shock might be oblique (Roettiger et al.~1995)\nocite{roettiger95}, or the
impact parameter of the merging clusters is non-zero (Briel et
al.~1991)\nocite{briel91}.

\begin{figure}[t]
\psfig{figure=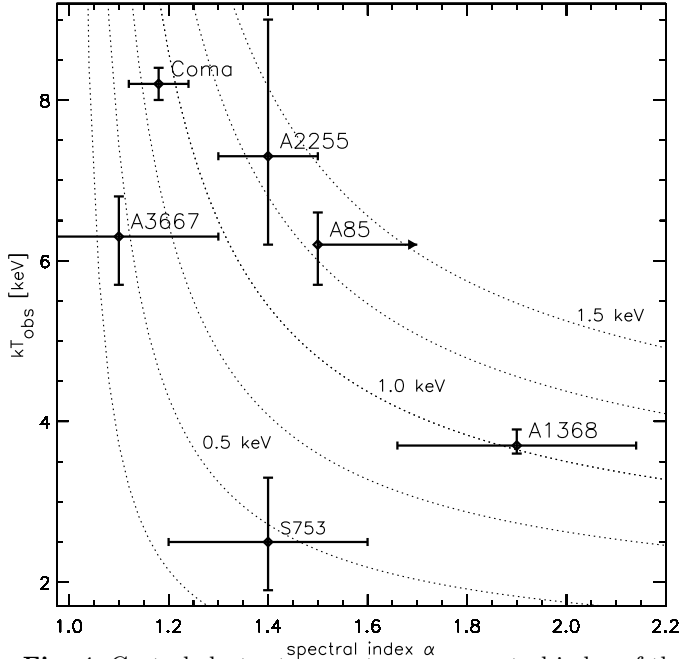,width=0.5\textwidth}
\caption[]{\label{Ts.ps}Central cluster temperatures vs. spectral
index of the radio relics. The dotted lines are temperature-spectral index
relations, computed from the temperature jump given by Eq. \ref{P}, which for a
given temperature of the infalling gas $kT_{1}$ (and assuming $kT_2=
\frac{1}{2} kT_{\rm obs}$) is a function of the spectral index $\alpha$ only.
The temperature of the infalling gas $kT_{1}$ is indicated for the curves.}
\end{figure}

\subsection{Abell 3667: 2006-56 { and 2010-57}}

The { giant} cluster radio relic 2006-56 is peripherally located within the
direction of the elongation of A3667. This elongation is seen in X-rays, but
also contains a subgroup of galaxies (Knopp et al. 1996). Because of the low
X-ray background, 2006-56 would be an ideal candidate for a future sensitive
search for inverse Compton flux (by, e.g., AXAF, XMM). Goss et
al.~(1982)\nocite{goss82} separated two extended radio sources at this
position, located close to each other: 2006-56 and a diffuse ridge.  { A
recent sensitive observation by R\"ottgering et
al.~(1997)\nocite{roettgering97} shows that there is in fact a single Z-shaped
structure, extending over $2.6\, {\rm Mpc}\, h_{50}^{-1}$, with constant
spectral index along the main axis of the relic. The Z-shape might be the
structure of the radio plasma, since we do not see the relic edge-on. The side
more distant from the cluster center is sharply edged and has a flatter
spectral index, as is expected to happen on the upstream side as explained in
Sect. \ref{sec:geometry}. The spectral index drops from $\alpha = 1.1$ in the
central region of the relic to $\approx 0.5$ at this edge.  This indicates
strongly the presence of particle acceleration, so that we can use the
thickness of the rim of lower spectral index in order to estimate the thickness
of the relic.  The projected thickness derived from the rim is approximately
$100- 200 \,{\rm kpc}\,h_{50}^{-1}$.  The region with steeper spectral
index on the inner side has also a projected thickness of $\approx 200 \,{\rm
kpc}\,h_{50}^{-1}$.  Assuming a viewing angle of $\delta= 48^\circ$,
deprojection gives $D\approx 150- 300\, {\rm kpc}\, h_{50}^{-1}$. The
projected area of the relic is $1.6\, {\rm Mpc}^2 \,h_{50}^{-2}$. A stripe of $
150\,{\rm kpc} \times 2.6\, {\rm Mpc}$ belongs to one edge. Deprojecting the
remaining area gives a relic surface of $1.8\, {\rm Mpc}^2 \,h_{50}^{-2}$. The
infalling gas seems to be relatively cool, with $kT_1 \approx 0.4$ keV, but
this number has a large error since the temperature jump is very sensitive to
small variations in $R$, since $R$ is close to its maximum value (see Eq.
\ref{P} or Fig. \ref{Ts.ps}).

R\"ottgering et al.  (1997)\nocite{roettgering97} report a pressure imbalance
of one to two orders of magnitude, between the gas pressure at the projected
cluster radius of the relic and its equipartition pressure. This discrepancy
should vanish if the lower gas pressure at the shock radius and the higher
radio plasma pressure of a compressed, flattened structure is taken into
account. Therefore we find good agreement between equipartition field strength
and predicted field strength.

Surprisingly, the radio image of R\"ottgering et al.
(1997)\nocite{roettgering97} reveals a second cluster radio relic on the other
side of the cluster (2010-57). These authors mention that both relics might be
the two lobes of a former, now inactive, radio galaxy, possibly the cD
galaxy. This must have been gigantic, since the projected distance is $5.2\,
{\rm Mpc}\, h_{50}^{-1}$. The expected diameter of the accretion shock sphere
of $8\, {\rm Mpc}\, h_{50}^{-1}$ would be a lower limit to the extent of this
radio galaxy in our scenario, since nowadays we would see a backflow of the
radio lobe from larger distances. In order to achive such an extent, the radio
galaxy must have been very powerful and the ambient medium more tenuous than
today. If both structures are really produced by one galaxy, this should have
happened during the early period of violent quasar activity, when the
intra-cluster medium was thinner.

If the two relics are independent, in the sense that their radio plasma is
from different sources, but their positions are correlated due to the
membership of A3667 to a large-scale filament, then the head-tail radio galaxy
B2007-569 can be the source of the shocked radio plasma in 2006-56. It is
located between the cluster center and the relic and is moving towards us with
a speed of 1200 km/s ($z=0.05257$; Sodre\'e et al.~1992).\nocite{sodree92} If
it is on a radially falling orbit, which traversed the shock sphere at the
location of the present day relic, it should have a velocity of 1800 km/s
(using $\delta=48^\circ$), higher than the expected infall velocity of matter
( measured in the cluster rest frame) at the higher radius of the
shock. Travelling with this velocity it should have passed the shock a Gyr
ago.  This is an order of magnitude longer than a typical cooling time of
radio emitting electrons and therefore the expected trace of radio plasma on
the path of that galaxy became invisible. It is amazing that R\"ottgering et
al.  (1997)\nocite{roettgering97} found marginal evidence for a bridge between
this galaxy and the relic, which might indicate reacceleration.}

{ We predict a polarization in 2006-56 of 28\% from our accretion shock
model, and 10\% in 2010-57 due to its smaller expected viewing angle of
$30^\circ$, assuming the same spectral index and shock compression ratio as
for 2006-56}. The observation of { these polarizations} would be an
independent test of the theory, and therefore could verify general assumptions
made about large-scale flows in the Universe.

\subsection{Abell S753: 1401-33}	

S753 is a poor cluster. Its low velocity dispersion of 536 km/s (Fadda et
al. 1996)\nocite{fadda96} indicates a low central gas temperature of $kT_{\rm
obs} = 2.5$ keV, applying the dispersion-temperature relation given in Bird et
al.~(1995)\nocite{bird95}, and therefore a flat gravitational potential
compared to the other clusters. The spectral index { $\alpha= 1.4$ of the
relic 1401-33 corresponds to a shock compression ratio of $R=2.7$}. This
relatively strong shock at this poor cluster can be understood if the
accreting matter has a temperature of $kT_{1}\approx 0.5$ keV. This is lower
than what seems to be typical for the other clusters ($kT_{1}\approx 1$ keV,
see Fig. \ref{Ts.ps}). The lower preheating of the flow onto this cluster
might indicate a poorer cosmological environment. Since the sky position of
the relic 1401-33 is close to the center of S753, the viewing angle should be
small and only 4\% polarization { is predicted}.

\section{Conclusions}

From simulations of the structure formation of the Universe, large accretion
shocks around clusters of galaxies are expected. If magnetic fields are present
at these locations, particle acceleration should take place.  Protons might be
accelerated to the highest observed energies, as demonstrated by Kang et
al. (1997), because of their weak energy losses { and their possible
injection out of a thermal population}. { If a population of relativistic
low-energy electrons is also present, they would be accelerated,
too. High-energy ultra-}relativistic electrons lose energy by inverse-Compton
scattering and synchrotron emission. This allows to see { them} in the radio
or X-ray regime and therefore verify the existence of { the large-scale
formation} shock structures.

At peripheral positions with respect to the cluster center regions of diffuse
radio emission are found, the so-called cluster radio relics ({
Figs. \ref{coma.ps} and \ref{skizze.ps}}). They are assumed to be remnants
of radio galaxies. Their energy supply was a long lasting outstanding problem,
since their electron cooling times are usually too short to allow any of the
nearby possible former parent galaxies to have moved from the relic to its
present position within that time. Electron acceleration has to take place
within them, but turbulent acceleration from galactic wakes fails to explain
this, due to the low galaxy densities in these regions.

If these relics are close to the accretion shock, a shock efficiency of only
$0.1-5\%$ is sufficient to power them.  The accretion shock itself should
collect relics since every volume filled with radio plasma, being injected
into the inter galactic medium above such a shock by a radio galaxy, would be
dragged into the shock by the fast motion of the infalling gas. After passing
the shock, the relic remains close to it for some time, since the growth rate
of the shock radius is much slower than the accretion velocity. Buoyant motion
of the light radio plasma embedded in the heavier surrounding gas might help
to keep it longer there.

The relic is proposed to be radio illuminated by relativistic electrons
(re)accelerated at the shock and therefore tracing its position. The radio
spectrum of such relics is steeper than expected for electron acceleration
within strong shocks, indicating that the temperature of the infalling matter
should be $\approx$ 0.5 - 1 keV { (see Fig. \ref{Ts.ps})}. Thus,
preheating of this gas should have taken place. { The accreting gas has a
typical density of $n_{\rm e,1}\approx 10^{-5}\, {\rm cm^{-3}}\,
h_{50}^{1/2}$. 1706+78 in A2256 of course does not fit into this temperature
scheme, due to its different physical nature, being located at a merger shock
front in the interior of the cluster. Matter which has fallen onto the
large-scale cosmic sheets should be shock-heated to about 0.1 keV and have a
density of $\leq 10^{-6}\, {\rm cm}^{-3}$ far away of clusters.  Adiabatic
compression and internal shocks of the converging flow within these sheets
onto clusters of galaxies should heat the gas further to the above
temperatures of $0.5- 1$ keV.}

During the passage through the shock, the magnetic fields are amplified and
aligned with the shock plane. The radio emission of a relic should therefore
exhibit polarization properties depending on the viewing angle.  Assuming a
spherical accretion shock with a radius depending on the depth of the cluster
potential, the viewing angle and the resulting polarization can be estimated
{ (Fig. \ref{pol:ps})}. These agree well with measured polarizations for
relics in two clusters: 1253+275 in Coma, and 1712+64 in A2255. For 0917+75 in
A786, the large projected radius would imply a deep gravitational potential
and therefore a very high cluster temperature, which is not yet
measured. 0917+75 might also trace an accretion shock of the stream onto the
supercluster Rood \#27, which A786 is a member of. This could also explain the
large apparent radius of the relic.

In case of 1706+78 in A2256 the measured polarization exceeds the expected one
significantly, indicating that the viewing angle is much higher than if the
relic were located at the accretion shock. A consistent shock radius would be 1
Mpc$\,h_{50}^{-1}$, which means that an internal shock of the cluster powers
the relic, which should result from a highly developed merger event. Roettiger
et al.~(1995)\nocite{roettiger95} discuss several apparent signs of such an
event in A2256, and fit simulations to the X-ray data.  Their best-fit value of
the angle between the merger axis and the line-of-sight within our theory of
field compression predicts exactly the observed polarization.

The directions of the projected magnetic fields are roughly perpendicular in
all above examples to the direction pointing to the cluster center, as is
expected.

For the other { five} relics of our sample no polarization measurements are
available. We predict $\approx 30\%$ polarization for 2006-56 in A3667, {
$\approx 10\%$ for 2010-57 in A3667}, and less than 5\% for 0038-096 in A85,
1140+203 in A1367, and 1401-33 in S753. Polarization measurements of these
sources could verify the existence of cluster accretion shocks and help us to
reveal an important piece of the cosmological structure formation puzzle.

The time elapsed since the magnetized plasma has started to pass the shock can
be calculated in two independent ways. Both estimates give ages of the order
of some $10^8$ yr, and are therefore a further justification of our model. 

{ Since our sample contains mostly relics seen under viewing angles below
$60^\circ$, we expect that more relics might be discovered in the outer
regions of clusters. Any radio search within these large areas should be guided
by the expected correlation of the large-scale structure filaments and typical
inflow directions into clusters.}

We conclude that the properties of cluster radio relics are naturally
explained if they are understood to trace some of the giant shock fronts of
the cosmological large-scale motion of the on-going structure formation. This
demonstrates for the first time the existence of these theoretically predicted
shocks. Estimated magnetic fields, temperatures and densities of the accreting
matter fit into the structure formation scenario.\\[0.3cm]

\noindent
{\it Acknowledgments.}  We would like to thank H.  Kang, D. Ryu for extensive
discussions on large-scale structure formation, Gopal Krishna for discussions
on radio relics, and the referee, L. Feretti, for useful comments and advice
which helped to improve the presentation of our work. We further acknowledge
G. Giovannini, L.  Feretti and C. Stanghellini for the permission to use their
figure. TAE acknowledges financial support by the {\it Studienstiftung
d. dt. Volkes}. This research has made use of the NASA/IPAC Extragalactic
Database (NED) which is operated by the Jet Propulsion Laboratory, California
Institute of Technology, under contract with the National Aeronautics and
Space Administration.

%

\newcounter{tabref}

 
\end{document}

%% file: relics.tab.tex
Cluster&
\scriptsize &
&
A85&
A786&
A1367&
Coma&
A2255&
A2256&
A3667&
S753\\

Relic&
\scriptsize &
{\scriptsize References}&
\scriptsize 0038-096&
\scriptsize 0917+75&
\scriptsize 1140+203&
\scriptsize 1253+275&
\scriptsize 1712+64&
\scriptsize 1706+78&
\scriptsize 2006-56&
\scriptsize 1401-33\\ \hline \hline 

$z$&
\scriptsize &
\scriptsize \cite{fadda96}\cite{harris93}\cite{zabludoff93}\cite{fadda96}\cite{fadda96}\cite{fadda96}\cite{fadda96}\cite{goss87}&
0.0559&
0.125&
0.0216&
0.0233&
0.0824&
0.0824&
0.0566&
0.0142\\

$kT_{\rm obs}$&
\scriptsize keV&
\scriptsize \cite{david95}{--}\cite{david93}\cite{briel92}\cite{david93}\cite{david93}\cite{knopp96}\cite{fadda96}&
6.2&
---&
3.7&
8.2&
7.3&
7.3&
6.3&
2.5$^{\scriptscriptstyle \rm (\ref{rem:kT})}$\\

$n_{\rm e,o}$&
\scriptsize $10^{-3}\,{\rm cm^{-3}}\,h_{50}^{1/2}$&
\scriptsize \cite{jones84}{--}\cite{jones84}\cite{briel92}\cite{feretti97}\cite{jones84}\cite{knopp96}{--}&
5.5&
---&
0.95&
3&
1.8&
2.5&
1&
---\\

$r_{\rm core}$&
\scriptsize kpc $h_{50}^{-1}$&
\scriptsize \cite{jones84}{--}\cite{jones84}\cite{briel92}\cite{feretti97}\cite{briel91}\cite{knopp96}{--}&
225&
---&
430&
400&
579&
352&
286&
---\\

$\beta$&
\scriptsize &
\scriptsize \cite{david95}{--}\cite{jones84}\cite{briel92}\cite{feretti97}\cite{briel91}\cite{knopp96}{--}&
0.62&
---&
0.52&
0.75&
0.74&
0.76&
0.54&
---\\

$r_{\rm projected}$&
\scriptsize Mpc $h_{50}^{-1}$&
\scriptsize \cite{lima97}\cite{harris93}\cite{gavazzi78}\cite{giovannini91}\cite{feretti97}\cite{roettgering94}\cite{goss82}\cite{goss87}&
0.64&
5&
1.09&
2.9&
1.25&
0.74&
2.95&
0.911\\

$\alpha$&
\scriptsize &
\scriptsize \cite{feretti96}\cite{harris93}\cite{gavazzi83}\cite{giovannini91}\cite{feretti97}\cite{roettgering94}\cite{roettgering97}\cite{goss87}&
$>$1.5$^{\scriptscriptstyle \rm (\ref{rem:A85})}$&
1.1&
1.9&
1.18&
1.4&
0.8$^{\scriptscriptstyle \rm (\ref{rem:s})}$&
1.1&
1.4\\

$P$&
\scriptsize \%&
\scriptsize {--}\cite{harris93}\cite{giovannini91}{--}\cite{feretti97}\cite{roettgering94}{--}{--}&
---&
54&
---&
27&
$<$2&
20&
---&
---\\

$B_{\rm 2,eq}$&
\scriptsize $\mu$G$\,h_{50}^{2/7}$&
\scriptsize \cite{feretti96}\cite{harris93}\cite{gavazzi83}\cite{giovannini91}\cite{feretti97}\cite{roettgering94}\cite{goss82}\cite{goss87}&
$<$1.4&
0.5&
2&
0.7&
0.6&
2.7&
1.6&
0.4\\

$D$&
\scriptsize Mpc $h_{50}^{-1}$&
\scriptsize \cite{lima97}\cite{harris93}\cite{gavazzi78}\cite{giovannini91}\cite{feretti97}\cite{roettgering94}\cite{goss82}\cite{goss87}&
$>$0.096&
0.22&
0.088&
0.1&
0.075&
0.1&
0.15$^{\scriptscriptstyle \rm (\ref{rem:deproj})}$&
0.082\\

$S$&
\scriptsize Mpc$^2$ $h_{50}^{-2}$&
\scriptsize \cite{lima97}\cite{harris93}\cite{gavazzi87}\cite{giovannini91}\cite{feretti97}\cite{roettgering94}\cite{roettgering97}\cite{goss87}&
0.096&
0.96&
0.053&
0.48&
0.19&
0.3&
1.8$^{\scriptscriptstyle \rm (\ref{rem:deproj})}$&
0.11\\

$\nu_{\rm break}$&
\scriptsize MHz&
\scriptsize \cite{joshi86}\cite{harris93}\cite{gavazzi83}\cite{giovannini91}\cite{feretti97}\cite{roettgering94}\cite{goss82}\cite{goss87}&
$\leq$300&
300&
$<$610&
$<$151&
$<$333&
$>$1500&
$<$85&
$<$843\\

$\zeta$&
\scriptsize &
\scriptsize \cite{joshi86}\cite{harris93}\cite{gavazzi83}\cite{giovannini91}\cite{feretti97}{--}\cite{goss82}\cite{goss87}&
$<$0.45&
$<$0.45&
$<$0.66&
$<$0.18&
$<$0.47&
---$^{\scriptscriptstyle \rm (\ref{rem:s})}$&
$<$0.13&
$<$0.75\\

$\log(Q_{\rm radio})$&
\scriptsize erg/s $h_{50}^{-2}$&
\scriptsize \cite{feretti96}\cite{harris93}\cite{gavazzi83}\cite{giovannini91}\cite{burns95}\cite{roettgering94}\cite{roettgering97}\cite{goss87}&
41.2&
41.6&
41.5&
40.9&
40.8&
41.5&
42.4&
41\\ \hline \hline 

$r_{\rm s}$&
\scriptsize Mpc $h_{50}^{-1}$&
\scriptsize Eq. \ref{eq:rs}&
4&
---&
3.2&
4.8&
4.1&
1.1$^{\scriptscriptstyle \rm (\ref{rem:merge})}$&
4&
2.6\\

$V_{\rm s}$&
\scriptsize km/s&
\scriptsize Eq. \ref{eq:Vs}&
1770&
---&
1370&
2040&
1920&
3000$^{\scriptscriptstyle \rm (\ref{rem:merge})}$&
1780&
1120\\

$\delta$&
\scriptsize $^\circ$&
\scriptsize Sect. \ref{compr}&
9.3$^\circ$&
---&
20$^\circ$&
37$^\circ$&
18$^\circ$&
42$^\circ$$^{\scriptscriptstyle \rm (\ref{rem:merge})}$&
48$^\circ$&
20$^\circ$\\

$R$&
\scriptsize &
\scriptsize Eq. \ref{rs}&
$<$2.5&
3.5&
2.1&
3.2&
2.7&
2.9$^{\scriptscriptstyle \rm (\ref{rem:s})}$&
3.5&
2.7\\

$P_2/P_1$&
\scriptsize &
\scriptsize Eq. \ref{P}&
$<$6&
26&
3.8&
15&
7.3&
9.3$^{\scriptscriptstyle \rm (\ref{rem:s})}$&
26&
7.3\\

$T_2/T_1$&
\scriptsize &
\scriptsize Eq. \ref{P}&
$<$2.4&
7.4&
1.8&
4.6&
2.7&
1.9$^{\scriptscriptstyle \rm (\ref{rem:s})}$&
7.4&
2.7\\


$n_{\rm e,1}$&
\scriptsize $10^{-5}\,{\rm cm^{-3}}\,h_{50}^{1/2}$&
\scriptsize Sect. \ref{Sec:Eff}&
$>$1.1&
---&
1.9&
0.35&
0.82&
6.2&
0.42&
---\\

$kT_1$&
\scriptsize keV&
\scriptsize Eq. \ref{P}&
$>$1.3&
---&
1&
0.88&
1.3&
3.8$^{\scriptscriptstyle \rm (\ref{rem:kTEA2256})}$&
0.42&
0.45\\

$V_{\rm s,predicted}$&
\scriptsize km/s&
\scriptsize Eq. \ref{eq:Vsp}&
1440&
---&
1020&
1850&
1600&
3050&
1690&
932\\

$B_{\rm 2,predicted}$&
\scriptsize $\mu$G$\,h_{50}^{1/4}$&
\scriptsize Sect. \ref{strongB}&
2.6&
---&
2.4&
1.9&
2.5&
10&
1.9&
---\\

$B_{\rm 1,eq}$&
\scriptsize $\mu$G$\,h_{50}^{2/7}$&
\scriptsize Sect. \ref{strongB}&
$>$0.57&
0.098&
1&
0.18&
0.22&
0.89&
0.31&
0.15\\

$B_{\rm 1,predicted}$&
\scriptsize $\mu$G$\,h_{50}^{1/4}$&
\scriptsize Sect. \ref{strongB}&
$>$1.1&
---&
1.2&
0.5&
0.94&
3.3&
0.38&
---\\

$\log(Q_{\rm flow})$&
\scriptsize erg/s $h_{50}^{-3/2}$&
\scriptsize Eq. \ref{eq:Qflow}&
$<$43.2&
---&
42.7&
43.7&
43.5&
45.2&
44.2&
---\\

$\eta_{\rm radio}$&
\scriptsize $\%\,h_{50}^{-1/2}$&
\scriptsize Sect. \ref{Sec:Eff}&
$>$0.99&
---&
6&
0.15&
0.18&
0.017&
1.3&
---\\



$t_{\rm age, kin}$&
\scriptsize $10^8$ year $h_{50}^{-1}$&
\scriptsize Sect. \ref{sec:diffusion}&
1.3&
---&
1.3&
1.5&
1&
0.98&
2.9&
1.9\\

$t_{\rm age, break}$&
\scriptsize $10^8$ year&
\scriptsize Eq. \ref{eq:tage}&
$\geq$3.2&
1.7&
$>$2.6&
$>$4&
$>$2&
$<$1.4&
$>$6.2&
$>$1.4\\

$\log(\kappa_2)$&
\scriptsize ${\rm cm^2/s}$ $h_{50}$&
\scriptsize Eq. \ref{eq:kappa2}&
$<\,$31&
$<\,$31.4&
$<\,$31.1&
$<\,$30.5&
$<\,$30.9&
---$^{\scriptscriptstyle \rm (\ref{rem:s})}$&
$<\,$30.5&
$<\,$30.9\\


$P_{\rm strong}$&
\scriptsize \%&
\scriptsize Eq. \ref{eq:Pstrong}&
0.92&
---&
4&
16&
3.4&
20&
28&
4.4\\


$\delta_{\rm strong}$&
\scriptsize $^\circ$&
\scriptsize Sect. \ref{strongB}&
---&
68$^\circ$&
---&
48$^\circ$&
$<$14$^\circ$&
43$^\circ$&
---&
---\\


$r_{\rm s,strong}$&
\scriptsize Mpc $h_{50}^{-1}$&
\scriptsize Sect. \ref{strongB}&
---&
5.4&
---&
3.9&
$>$5.3&
1.1&
---&
---\\
 \hline 